\documentstyle[aps,preprint,epsfig]{revtex}
\begin{document}
\preprint{\vbox{\hbox{UMD PP\#01-030}\hbox{DOE/ER/40762-216}}}

\title{Counting Form Factors of Twist-Two Operators} 

\author{Xiangdong Ji\footnote{xji@physics.umd.edu}}
\address{Department of Physics, 
University of Maryland, College Park, Maryland 20742 }

\vskip 0.1in

\author{Richard F. Lebed\footnote{Richard.Lebed@asu.edu}}
\address{Department of Physics and Astronomy,
Arizona State University, Tempe, Arizona 85287 }

\vskip 0.1in
\date{December, 2000}
\vskip 0.1in

\maketitle
\tightenlines
\thispagestyle{empty}

\begin{abstract}
We present a simple method to count the number of hadronic form
factors based on the partial wave formalism and crossing symmetry. In
particular, we show that the number of independent nucleon form
factors of spin-$n$, twist-2 operators (the vector current and
energy-momentum tensor being special examples) is $n+1$.  These
generalized form factors define the generalized (off-forward) parton
distributions that have been studied extensively in the recent
literature.  In proving this result, we also show how the $J^{PC}$
rules for onium states arise in the helicity formalism.
\end{abstract}


\setcounter{page}{1}

The generalized (off-forward) parton distributions of hadrons, the
nucleon in particular, have attracted considerable theoretical and
experimental interest in the last few years \cite{ji1}. These
distributions generalize the well-known Feynman parton distributions
as well as the elastic electromagnetic form factors. Most
interestingly, the distributions contain information about the orbital
motion of partons in a polarized nucleon. One can, for instance,
deduce the amount of the nucleon spin carried by the quark orbital
angular momentum once these distributions are known
\cite{ji2}.

One way to define the generalized parton distributions is to consider
the nucleon matrix elements of the twist-2 operators. An example is
the chiral-even spin-independent quark operators
\begin{equation}
     \hat O^{\mu_1\cdots\mu_n} = \overline{\psi}(0) \, i \!
     \stackrel{\leftrightarrow}{D}^{(\mu_1} \! \! \cdots i \!
     \stackrel{\leftrightarrow}{D}^{\mu_n)} \! \psi(0) ~~~ n=1,2,...\
     ,
\end{equation}
where $\stackrel{\leftrightarrow}{D}^{\mu_1}=
(\stackrel{\leftarrow}{D}^{\mu_1} \! \! -\stackrel{\rightarrow}
{D}^{\mu_1})/2$, which form totally symmetric tensor representations
of the Lorentz group when the Lorentz indices $\mu_1,\cdots,\mu_n$ are
symmetrized and rendered traceless (indicated by parentheses).
Clearly, this tower of operators is a generalization of familiar
vector current $\bar\psi\gamma^\mu\psi$. The forward matrix elements
of the above operators define the {\it generalized charges}, $a_n$
\begin{equation}
    \langle P|\hat O^{\mu_1\cdots\mu_n}|P\rangle
  = 2a_n(Q^2) P^{\mu_1}\cdots P^{\mu_n} \ , 
\end{equation}
where $Q^2$ is a renormalization scale. The nucleon state is
normalized covariantly: $\langle P^\prime|P\rangle = (2P^0)(2\pi)^3
\delta^3(\vec{P}-\vec{P}^\prime)$. Feynman's unpolarized quark
distribution $q(x,Q^2)$ can be obtained directly from the generalized
charges,
\begin{equation}
  \int^1_{-1} x^{n-1} q(x, Q^2) dx = a_n(Q^2) \ , 
\end{equation}
with $n\ge 1$. As usual, $q(x, Q^2)$ at negative $x$ represents the
distribution of antiquarks.
     
The off-forward matrix elements of the twist-2 operators define {\it
generalized form factors} $A_{n,2i}(t, Q^2)$, $B_{n,2i}(t, Q^2)$ and
$C_n(t, Q^2)$ \cite{ji1},
\begin{eqnarray}
\langle P^\prime| \hat O^{\mu_1\cdots \mu_n} |P\rangle
   &= & \ \ \ \ {\overline U}(P^\prime) \gamma^{(\mu_1} U(P)
\sum_{i=0}^{[{n-1\over 2}]}
       A_{n,2i}(t,Q^2) \, q^{\mu_2}\cdots q^{\mu_{2i+1}} 
      \overline{P}^{\mu_{2i+2}}\cdots
      \overline{P}^{\mu_n)}  \nonumber \\ 
    &&  + ~  {\overline U}(P^\prime){\sigma^{(\mu_1\alpha} 
     iq_\alpha \over 2M}U(P)   \sum_{i=0}^{[{n-1\over 2}]}
     B_{n,2i}(t, Q^2) \, q^{\mu_2}\cdots q^{\mu_{2i+1}} 
      \overline{P}^{\mu_{2i+2}}\cdots
    \overline{P}^{\mu_n)} \nonumber \\
     &&  + ~{\rm mod}(n+1,2)~{1\over M}\bar U(P^\prime) U(P) ~ C_{n}(t, Q^2)
    \, q^{(\mu_1} \cdots q^{\mu_n)} \ , 
\label{form}
\end{eqnarray}
where ${\overline U}(P^\prime)$ and $U(P)$ are the Dirac spinors and
mod$(n+1,2)$ is 1 when $n$ even, 0 when $n$ odd.  The four-momentum
transfer is denoted by $q = P^\prime-P$, $t=q^2$, and $\overline P =
(P^\prime+P)/2$.  For $n\ge 1$, even or odd, there are exactly $n+1$
form factors. The moments of the off-forward parton distributions
$H(x,\xi,t)$ and $E(x,\xi,t)$ \cite{ji1} are
related to linear combinations of the above form factors.

The key question we address in this short paper is why there are
exactly $n+1$ form factors.  We anticipate that the answer also
addresses the general question of how to count the number of form
factors of any operator between hadron states.

To motivate a simple counting procedure, we first count the helicity
amplitudes in the Breit frame.  We start with the simplest example:
the vector current $\hat O^\mu = \bar \psi\gamma^\mu\psi$, which is
known to define two form factors for the nucleon states,
\begin{equation}
   \langle P^\prime|\hat O^\mu|P\rangle 
   = \overline{U}(P^\prime)\left[F_1(q^2)\gamma^\mu + 
   F_2(q^2) {i\sigma^{\mu\nu}q_\nu\over 2M}\right]U(P) \ . 
\end{equation}
To understand why there are two, consider the Breit frame, in which
the initial and final nucleon have 3-momenta of the same magnitude but
opposite directions. The 4-vector current can be decomposed into a
3-vector source plus a 3-scalar source.  Because of current
conservation, the scalar degree of freedom does not provide independent
information.  Thus, one need only count the number of independent
helicity amplitudes for $P + \mbox{\rm (3-vector~source)}\rightarrow
P^\prime$. Because of the collinearity of the process in this frame,
the angular momentum projection along the direction of the reaction is
conserved. Hence, there are four possible helicity amplitudes:
$A_{1/2, \, 0 \rightarrow -1/2}$, $A_{-1/2, \, 0 \rightarrow 1/2}$,
$A_{1/2, \, 1\rightarrow 1/2}$, and $A_{-1/2,\, -1\rightarrow -1/2}$,
where the indices label the helicities of the initial nucleon, the
source, and the final nucleon, respectively. However, amplitudes that
differ only by flipping the signs of all helicities are related
through parity invariance, and therefore only two of the amplitudes
are independent, as expected.

A subtlety arises if the vector current is not conserved.  Then one
has an extra spin-0 source that leads to a new helicity amplitude
$A_{1/2, 0^\prime \rightarrow -1/2}$ (and its parity partner),
suggesting a total of three independent form factors.  It turns out,
however, that if the (hermitian) vector current transforms like $\bar
\psi\gamma^\mu\psi$ under time reversal, the new amplitude vanishes
due to time-reversal symmetry. Indeed, let us write down possible
invariant form factors in this case,
\begin{equation}
   \langle P^\prime|\hat O^\mu|P\rangle 
   = \overline{U}(P^\prime)\left[F_1(q^2)\gamma^\mu + 
   F_2(q^2) {i\sigma^{\mu\nu}q_\nu\over 2M}
  + F_3(q^2)iq^\mu \right]U(P) 
   \ , 
\end{equation}
where according to hermiticity we have included a factor of $i$ in
front of $q^\mu$ so that $F_3(q^2)$ is real: Since hermitian
conjugation exchanges bra and ket, then $P \leftrightarrow P^\prime$
in the external states, and the sign of $q = P-P^\prime$ is flipped
(Note that this requirement only holds if the initial and final
particles are the same).  On the other hand, time reversal symmetry
requires
\begin{eqnarray}
  \langle P^\prime|\hat O^\mu|P\rangle 
  &=& \langle P^\prime|T^{-1}T \hat O^\mu T^{-1}T|P\rangle^* 
 \nonumber \\
  &=& \overline{U}(P^\prime)\left[F_1(q^2)\gamma^\mu + 
   F_2(q^2) {i\sigma^{\mu\nu}q_\nu\over 2M}
  - F_3(q^2)iq^\mu\right]U(P) \ ,
\end{eqnarray}
where in the second line we use $T\hat O^\mu T^{-1} = \hat O_\mu$ and
the time-reversal transformation of the states.  Therefore $F_3(q^2)$
must vanish, and we again only have two form factors.

The above example shows that helicity counting in the Breit frame
cannot provide the correct answer unless time-reversal symmetry is
taken into account.  This is easy to do for helicity amplitudes in
elastic scattering.  For form factors, however, the form of the
constraint is less clear.  One might follow the above approach by
writing down all possible invariant form factors and imposing
time-reversal symmetry: One then obtains Eq.~(\ref{form}).
Unfortunately, this procedure does not provide a simple way to
understand the physics underlying the counting.

In the remainder of the paper, we present an alternative counting
method. We first recall a basic property of relativistic quantum field
theory that the number of independent amplitudes is the same in all
crossed channels. Then we count the number of independent matrix
elements, $\langle P\overline{P}|\hat O^{\mu_1\cdots\mu_n}| 0\rangle$,
corresponding to $P\overline{P}$ creation from the twist-2 source. In
the center of momentum (c.m.)  frame, the enumeration of possible
$P\overline{P}$ states is well-known: Since $S=0, 1$,
$\vec{J}=\vec{L}+\vec{S}$, $P=(-1)^{L+1}$, and $C=(-1)^{L+S}$, the
list of allowed $J^{PC}(L)$ reads
\begin{eqnarray}
&&0^{++}(1), ~~0^{-+}(0), \nonumber \\
&&1^{++}(1), ~~1^{+-}(1), ~~1^{--}(0), ~~1^{--}(2), \nonumber \\
&&2^{++}(1), ~~2^{++}(3), ~~2^{-+}(2), ~~2^{--}(2), \nonumber \\
&&3^{++}(3), ~~3^{+-}(3), ~~3^{--}(2), ~~3^{--}(4), \nonumber  \\
&&\cdots .
\label{jpc}
\end{eqnarray}
For each $J\ge 1$, there are four possible states. Two of them have
the same $J^{PC}$ but different $L$: $J^{(-1)^J, \, (-1)^J}$, with
$L=J \pm 1$.

One might suspect that the classification of $P\overline{P}$ into
states of definite $J^{PC}$ using $\vec{L}$ (orbital angular momentum
being defined in the $P\overline{P}$ c.m.\ frame) and $\vec{S}$ (each
nucleon spin being defined in its own rest frame) is inherently
non-relativistic.  This is not so because the relativistic spin wave
functions transform the same way under space rotations as their
non-relativistic counterparts~\cite{weinberg}. The fully relativistic
treatment~\cite{landau} employing the helicity formalism gives exactly
the same rules for allowed $J^{PC}$ with the same multiplicity of
amplitudes, as we now show.

	Let $\psi_{JM \lambda_1 \lambda_2}$ be a two-particle helicity
state.  With $\eta_{1,2}$ being the intrinsic parities of the two
particles, the action of parity gives
\begin{equation}
\hat P \psi_{JM \lambda_1 \lambda_2} = \eta_1 \eta_2 (-1)^J
\psi_{JM -\lambda_1 -\lambda_2} \ ,
\end{equation}
a property we used above in reducing the number of helicity
amplitudes.  The factor $\eta_1 \eta_2$ is unity for any
particle-antiparticle pair.  Similarly, the action of charge
conjugation on such a self-conjugate pair gives
\begin{equation}
\hat C \psi_{JM \lambda_1 \lambda_2} = (-1)^J \psi_{JM \lambda_2
\lambda_1} \ ,
\end{equation}
Thus, suppressing the $JM$ subscripts, one may form states of definite
$J^{PC}$:
\begin{eqnarray}
PC=++:~~~ & & \psi_{\lambda_1 \lambda_2} + (-1)^J \psi_{\lambda_2
\lambda_1} + (-1)^J \psi_{-\lambda_1 -\lambda_2} + \psi_{-\lambda_2
-\lambda_1} \ , \nonumber \\
PC=+-:~~~ & & \psi_{\lambda_1 \lambda_2} - (-1)^J \psi_{\lambda_2
\lambda_1} + (-1)^J \psi_{-\lambda_1 -\lambda_2} - \psi_{-\lambda_2
-\lambda_1} \ , \nonumber \\
PC=-+:~~~ & & \psi_{\lambda_1 \lambda_2} + (-1)^J \psi_{\lambda_2
\lambda_1} - (-1)^J \psi_{-\lambda_1 -\lambda_2} - \psi_{-\lambda_2
-\lambda_1} \ , \nonumber \\
PC=--:~~~ & & \psi_{\lambda_1 \lambda_2} - (-1)^J \psi_{\lambda_2
\lambda_1} - (-1)^J \psi_{-\lambda_1 -\lambda_2} + \psi_{-\lambda_2
-\lambda_1} \ . \label{hel}
\end{eqnarray}
For spin-1/2 particles, the only independent choices are $\lambda_1 =
\lambda_2$ and $\lambda_1 = -\lambda_2$.  In the former case,
Eqs.~(\ref{hel}) show that only the two amplitudes with $C=(-1)^J$
(and either parity) are nonvanishing, while in the latter case only
the two amplitudes with $PC=+1$ survive.  These are the allowed
sequences $0^{++}, 0^{-+}, 1^{+-}, 1^{--}, 2^{++}, 2^{-+}, \cdots$,
and $1^{++}, 1^{--}, 2^{++}, 2^{--}, \cdots$, respectively.  Note that
the amplitudes $0^{++}$ and $0^{--}$ have not been included in the
second list, since their amplitudes according to Eq.~(\ref{hel}) would
be $\psi_{\lambda -\lambda} \pm \psi_{-\lambda \lambda}$; however,
since the particles are back-to-back in the c.m., the existence of
each term requires an angular momentum projection along the axis of
$+1$ or $-1$, which is forbidden since $J=0$.  Hence, the $0^{++}$
amplitude with $\lambda_1 = -\lambda_2$ and the $0^{--}$ amplitude
vanish.  Thus, the complete set of forbidden states (the so-called
``exotics'' in the context of the quark model) is, precisely as in the
non-relativistic counting, $0^{--}, 0^{+-}, 1^{-+}, \cdots, J^{(-1)^J,
\, (-1)^{J+1}}, \cdots$.  Note also that the amplitudes $J^{(-1)^J, \,
(-1)^J}$ appear twice in the ``non-exotic'' list (except $0^{++}$,
which only appears once), exactly as in the non-relativistic case.

Clearly, only an external source with the same $J^{PC}$ can produce
these states. Let us classify $J^{PC}$ of the twist-2 operators. $\hat
O^{\mu_1\cdots\mu_n}$ furnishes an $(n/2,n/2)$ representation of
Lorentz group \cite{weinberg} and has $(n+1)^2$ independent components: Under
spatial rotations, the tensor is decomposed into the angular momentum
components $J^P=0^+, 1^-, \cdots, n^{(-1)^n}$, each with natural
parity $(-1)^J$. The charge-conjugation parity is clearly $(-1)^n$, a
sign appearing for each of the $n$ Lorentz indices.  Thus, the
$J^{PC}$ content of the twist-2 operator is
\begin{eqnarray}
&&n=0, ~~0^{++}, \nonumber \\
&&n=1, ~~0^{+-}, ~~1^{--}, \nonumber \\
&&n=2, ~~0^{++}, ~~1^{-+}, ~~2^{++}, \nonumber \\
&&n=3, ~~0^{+-}, ~~1^{--}, ~~2^{+-}, ~~3^{--}, \nonumber \\
&&\cdots .
\end{eqnarray}
Now counting the number of independent matrix elements is
straightforward.  Since the use of $L$ may be more familiar to the
reader, and gives the same counting as the helicity formalism, we
retain the $L$ labels for convenience.  For $n=1$, only the $1^{--}$
source is effective. According to Eqs.~(\ref{jpc}), it can create two
$1^{--}$ states (with $L=0$ and $2$), and hence there are two
independent matrix elements. For $n=2$, both $0^{++}$ and $2^{++}$
sources are effective. While $0^{++}$ can only create one state, the
$2^{++}$ source can create two independent states (with $L=1$ and
3). Therefore, there are three independent matrix elements.  A list of
matrix elements in terms of the quantum numbers reads
\begin{eqnarray}
&& n=0, ~~ J^{PC}(L) = 0^{++}(1),  \nonumber \\
&& n=1, ~~ J^{PC}(L) = 1^{--}(0), ~~1^{--}(2),   \nonumber  \\
&& n=2, ~~ J^{PC}(L) = 0^{++}(1), ~~2^{++}(1), ~~2^{++}(3),   \nonumber \\
&& n=3, ~~ J^{PC}(L) = 1^{--}(0), ~~1^{--}(2), ~~3^{--}(2), ~~3^{--}(4),
\nonumber \\ && \cdots . 
\end{eqnarray}
This simple pattern is easily extended and proves that there are $n+1$
independent matrix elements for arbitrary $n\ge 0$, implying $n+1$
form factors in $\langle P^\prime|\hat O^{\mu_1\cdots
\mu_n}|P\rangle$.

The method presented above is completely general. One can use it to
count the form factors of a general operator in any hadron states.
The procedure is summarized here again: First one goes to the crossed
channel in which the operator serves as a source for creating a
particle-antiparticle pair. Then allowed $J^{PC}$ values for the
operator and particle-antiparticle pair are enumerated and matched.
Crossing symmetry plus $C$ and $P$ invariance replaces the use of $T$
invariance in the direct channel.  Finally, for each $J^{PC}$, the
number of form factors is determined by the number of independent
amplitudes for the creation process. For example, it is easy to show
by the same technique that the number of independent form factors of
the twist-2 operators in a spinless state like the pion is $[n/2]+1$:
Here, Eq.~(\ref{hel}) shows that only the series $0^{++}, 1^{--},
2^{++}, \cdots$ occurs, each of these amplitudes appearing only once.

To summarize, we have presented a simple and general method to count
the number of form factors. The method provides a straightforward
verification that the number of form factors of a twist-2, spin-$n$
operator is $n+1$ for nucleon states.

\acknowledgments
We wish to acknowledge the support of the U.S.~Department of Energy
under Grant Nos.\ DE-FG02-93ER-40762 (X.J.) and DE-AC05-84ER40150
(R.F.L.).  R.F.L. also thanks the U. Maryland TQHN group for their
hospitality.

\end{document}